\documentclass[useAMS,usenatbib,usegraphicx]{mn2e}
\usepackage{txfonts}
\title[SXP18.3 undergoes the longest Type II outburst in the SMC]{High mass X-ray binary SXP18.3 undergoes the longest Type II outburst ever seen in the Small Magellanic Cloud}
\author[M.P.E. Schurch et al.]{M.P.E. Schurch$^{1}$\thanks{E-mail:
mpes@astro.soton.ac.uk},  M.J. Coe$^{1}$, J.L. Galache$^{2}$, R.H.D. Corbet$^{3}$, K.E. McGowan$^{1}$, 
\newauthor V.A. McBride$^{1}$, L.J. Townsend$^{1}$, A. Udalski$^{4}$, F. Haberl$^{5}$\\
$^{1}$School of Physics and Astronomy, Southampton University, Highfield, Southampton, SO17 1BJ, UK.\\
$^{2}$Harvard-Smithsonian Center for Astrophysics, 60 Garden Street, Cambridge, MA 02138, USA.\\
$^{3}$University of Maryland Baltimore County, X-ray Astrophysics Laboratory, Mail Code 662, NASA Goddard Space Flight Center, Greenbelt, MD 20771, USA.\\
$^{4}$Warsaw University Observatory, Aleje Ujazdowskie 4, 00-478 Warsaw,
Poland.\\
$^{5}$Max-Planck-Institut f\"ur extraterrestrische Physik, Giessenbachstra\ss{}e, 85748, Germany.}

\begin{document}

\date{Accepted for publication - 06 October 2008}

\pagerange{\pageref{firstpage}--\pageref{lastpage}} \pubyear{2008}

\maketitle

\label{firstpage}

\begin{abstract}
On 30th August 2006 SXP18.3 a high mass X-ray binary (HMXB) in the Small Magellanic Cloud (SMC) with an 18.3\,s pulse period was observed by {\it Rossi X-ray Timing Explorer} ({\it RXTE}).  The source was seen continuously for the following 36 weeks.  This is the longest Type II outburst ever seen from a HMXB in the SMC.  During the outburst SXP18.3 was located from serendipitous {\it XMM-Newton} observations.  The identification of the optical counterpart has allowed SXP18.3 to be classified as a Be/X-ray binary.  This paper will report on the analysis of the optical and weekly {\it RXTE} X-ray data that span the last 10 years.  The extreme length of this outburst has for the first time enabled us to perform an extensive study of the pulse timing of an SMC Be/X-ray binary.  We present a possible full orbital solution from the pulse timing data.  An orbital period of 17.79\,d is proposed from the analysis of the OGLE III light curve placing SXP18.3 on the boundary of known sources in the Corbet diagram. 
\end{abstract}


\section{Introduction}
Over the past 20 years the number of known high-mass X-ray binaries (HMXBs) located in the Small Magellanic Cloud (SMC) has been steadily increasing.  There are now 58 known HMXB pulsar systems in the SMC \citep{coe05}. Systems where the counterpart is a Be star form the largest sub-class of HMXBs.  Within the SMC all but one system (SMC X-1) are found to have Be star counterparts.  \citet{mcbride08} find that the spectral distribution of these counterparts is consistent with the distribution found in the Galaxy.  From studies of Be/X-ray binaries in the Galaxy it has been found that the compact object, presumably a neutron star is generally in a wide and eccentric orbit.  X-ray outbursts are normally associated with the passage of the neutron star close to the circumstellar disk \citep{oka01}.  Detailed reviews of the X-ray and optical properties of such systems may be found in \citet{coe05} and \citet{haberl04}, a review of the more general properties can be found in \citet{coe00}.

\subsection{SXP18.3 = XTE J0055-727 = XMMU J004911.4-724939}
SXP18.3 was discovered in 2003 during routine {\it Rossi X-ray Timing Explorer} ({\it RXTE}) proportional counter array (PCA) observations of the SMC \citep{cor03}.  They report detecting pulsations with a period of $18.37\pm0.1s$.  {\it RXTE} slew observations were able to narrow the position down to an ellipse with R.A. = $13.84^\circ\pm0.1$ and DEC. = $72.70^\circ\pm0.06$, however, no firm identification of an optical counterpart was possible.  Five detections of SXP18.3 were made in 2004.  \citet{cor04} noticed that these detections appeared to be occurring on a 34.8\,d period and proposed this as the orbital solution.  From the full analysis of the X-ray light curve prior to the large outburst \citet{gal08} refined this period to $17.37\pm0.01\,d$ \citep[approximately half of the value proposed by][]{cor04}. On 12-13 March 2007 SXP18.3 was identified serendipitously through a 39ks {\it XMM-Newton} observation of the emission nebula N19 in the SMC \citep{eger08,haberl08}.  They report the presence of a bright X-ray transient source showing pulsations at $18.3814\pm0.0001s$, at R.A. = 00:49:11.4, DEC = -72:49:39 with a positional uncertainty of 1.18".  The detection of these pulsations confirms the identification as being SXP18.3 and enabled the optical counterpart to be identified. In this paper we present the analysis of both the long 10\,year optical and X-ray light curves, as well as constraints on the orbital parameters derived from detailed timing analysis of the detected pulse period.

\begin{table*}
 \centering
  \caption{Optical counterpart matches. "Distance" refers to the offset of the object from the {\it XMM-Newton} position. \label{ta:ids}}
  \begin{tabular}{@{}llrrrrrrrr@{}}
  \hline
   Catalogue & Reference ID & RA & DEC & Distance (") & Date (MJD) & U & B & V & I\\
  \hline
   MCPS$^{1}$ & 1933826 & 00:49:11.53 & -72:49:37.2 & 0.11 & - & $15.22\pm0.03$ & $16.01\pm0.02$ & $15.96\pm0.03$ & $15.89\pm0.04$\\
   MACHO$^{2}$ & 208.15911.13 & 00:49:11.42 & -72:49:36.1 & 1.31 & 48855-51527 & - & - & 16.0-15.2 & 16.0-15.4 \\
  \hline  
   Catalogue & Reference ID & RA & DEC & Distance (") & Date (MJD) & I & J & H & K$_{s}$\\
  \hline
   OGLE II$^{3}$ & smc\_ sc5 65500 & 00:49:11.45 & -72:49:37.1 & 0.45 & 50466-51871 & 16.0-15.2 & - & - & -\\
   OGLE III$^{4}$ & SMC101.8 19552 & 00:49:11.45 & -72:49:37.1 & 0.45 & 52086-54487 & 15.9-14.8 & - & - & -\\
   2MASS$^{5}$ & 00491147-7249375 & 00:49:11.47 & -72:49:37.6 & 0.43 & 51034 & - & $15.9\pm0.1$ & $>15.119$ & $15.6\pm0.2$\\
   SIRIUS$^{6}$ & 00491143-7249375 & 00:49:11.43 & -72:49:37.5 & 0.52 & 52517 & - & $16.05\pm0.02$ & $16.03\pm0.03$ & $16.13\pm0.05$\\
  \hline
  \end{tabular}\\ 
 \begin{flushleft}
   $^{1}$\citet{zar02}, $^{2}$\citet{alc99}, $^{3}$\citet{uda97,szy05}, $^{4}$ Private communication, $^{5}$\citet{skr06}, $^{6}$\citet{kato07}.
\end{flushleft}
\end{table*}  

\section{Observations and Data Analysis}
\subsection{Optical Observations}
The optical counterpart is clearly identified by the {\it XMM-Newton} position by \citet{eger08} as a V=16\,mag star \citep{zar02}. This counterpart is present in a number of optical catalogues of the SMC.  These data are summarised in Table \ref{ta:ids}, where we also quote the positional offset from the {\it XMM-Newton} position.  The optical magnitudes both in the visual range and the infra-red are typical of the many Be star counterparts that are associated with the other Be/X-ray binares in the SMC \citep{coe05,schurch07}.

\begin{figure}
 \includegraphics[width=60mm, angle=270]{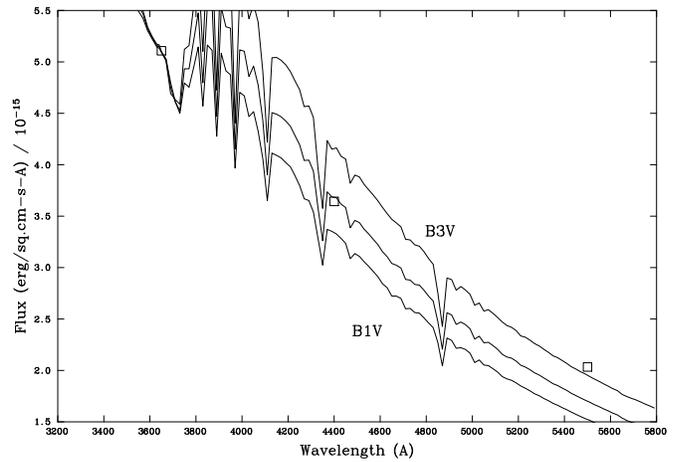}
  \caption{Comparison of U,B, and V MCPS fluxes, dereddened by E(B-V)=0.09, with Kurucz stellar atmosphere models.\label{fig:o_spectra}}
\end{figure}

To compare the MCPS photometry (Table \ref{ta:ids}) with standard stellar models \citep{kurucz79} the data were first corrected for the standard reddening to the SMC of E(B-V)=0.09 \citep{schwering91}.  The dereddened MCPS colour (B-V)=-0.04 is typical of the Be star counterparts found in the SMC \citep{mcbride08,shty05}.  This colour suggests an optical counterpart in the range B0-B2.  Figure \ref{fig:o_spectra} shows the dereddened MCPS U,B, and V  fluxes plotted over possible Kurucz stellar atmosphere models.  The U and B fluxes suggest that the counterpart is in fact B2V, whilst the V flux indicates that there is a red excess typical of a circumstellar disk.  This classification will need verification from detailed spectrographic measurements. However, in light of this fit we support the proposal by \citet{eger08} that SXP18.3 is a Be/X-ray binary.

\subsubsection{Optical light curves}
\begin{figure}
 \includegraphics[width=60mm, angle=90]{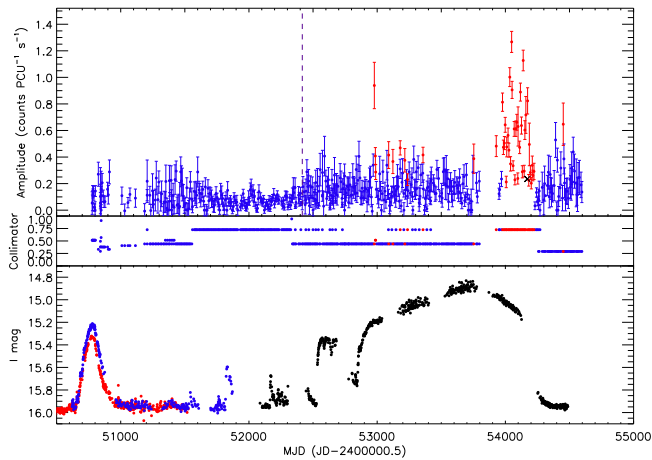}
  \caption{SXP18.3 X-ray and optical light curves. Panels are PCA pulsed flux light curve in the energy range 3-10keV (top), PCA collimator response (middle), combined MACHO (red), OGLE II (blue) and III (black) light curves (bottom). The black cross represents the {\it XMM-Newton} detection (MJD 54171) and the dashed line indicates a non detection by {\it Chandra} ACIS-I, observation 2944 (MJD 52416). The red points in the {\it RXTE} light curve and collimator response represent detections of the source above a 99\% local significance level.\label{fig:xlc}}
\end{figure}

\begin{figure}
\includegraphics[width=60mm, angle=90]{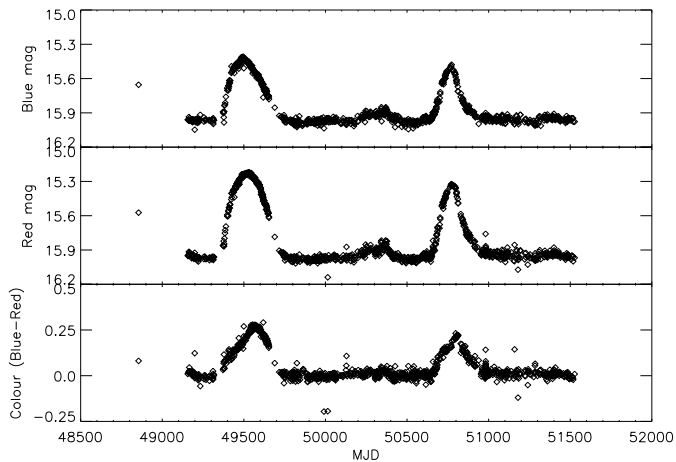}
\caption{MACHO red and blue light curves and colour variations.\label{fig:mcolour}}
\end{figure}

From the combined OGLE II and III light curves we scaled the instrumental MACHO magnitudes so that the continuum of both the OGLE II and MACHO data align.  Shifts of +24.81 and +24.55 magnitudes were applied to the blue and red MACHO data respectively.  The bottom panel of Figure \ref{fig:xlc} shows the combined light curve from August 1992 to April 2008 (only the overlapping MACHO data are shown).  The full MACHO red and blue lightcurves are shown in Figure \ref{fig:mcolour}.  As is clearly seen there have been a number of large distinct optical outbursts.  The first two lasted for 500-600\,d each, and later one giant outburst started around MJD\,52800 that lasted for around 1500\,d.  After each outburst the optical flux has always returned to an approximately steady base value of $\sim$16.0-15.9\,mag.
The simultaneous red and blue MACHO data allows the colour changes throughout the first two optical outbursts to be examined.  It is clear that as each of the outbursts proceeded the source slowly reddens.  After the peak of the outburst is reached the colour returns to its baseline configuration.  It is worth noting that the colour change is asymmetrical.  Similar behaviour is also observed in SXP6.85 \citet{kem08}.  As was suggested in \citet{kem08} the colour variation would likely indicate changes in the structure of the circumstellar disks.  Either the formation of the disk is increasing the optical brightness by the addition of red light or the large disk is now masking out part of the surface of the bluer Be star, behaviour that is dependant on the inclination angle of the system.

\subsubsection{Orbital Period}
The large optical outbursts seen in the light curve are too bright and occur on vastly longer time scales than would be expected for this source if they were due to the orbital period ($\sim$20-100 days from the Corbet diagram).  We attribute them to changes in the size of the circumstellar disk.  However, a search for possible binary modulations was carried out on the detrended data.  Removing the large variations without introducing artificial variations and features proved to be extremely difficult.  Periods when the light curve is changing very fast are impossible to detrend meaningfully.  We thus restricted the search to the last 4 years of OGLE III data (MJD 53164.9 to the 54487.6) where we have been able to detrend the data by subtracting linear fits.  A single linear fit was used for the first three years.  The fourth year of data was first split in half and then a separate linear fits were used for each half. The resultant light curve was searched as both a whole and as individual years for periodicities in the range 1-100\,d using Lomb-Scargle (LS) periodograms. 

\begin{figure}
 \includegraphics[width=60mm, angle=90]{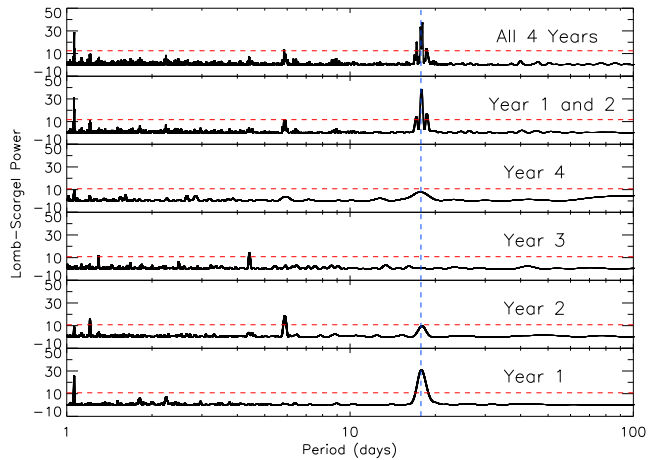}
  \caption{Power spectra of the 4 year detrended OGLE III light curve. The dotted horizontal red lines indicate the 99\% significance level and the dotted vertical blue line indicates the 17.79\,d period.\label{fig:o3ps}}
\end{figure}

The results of the temporal analysis are shown in Figure \ref{fig:o3ps}.  A clear peak well above the 99\% significance level occurs at $17.79\pm0.01$d in the first two years of our detrended OGLE III data, in agreement with \citet{uda08}.  This periodicity then disappears completely during the third year only to reappear at a much lower significance in the 4th year at a slightly different period. The lobes in the combined year 1 and 2 data are due to the 17.79\,d period beating with the one year sampling.  The very short period peaks (around 1\,d) are due to the one day sampling of the data. We also note the high peaks at 5.87\,d in the year 2 data and the peak at 4.40\,d in the year 3 data.  These peaks are not the third and fourth harmonics (respectively) of the higher period peak at 17.79\,d.  As a consequence of its strength the 5.87\,d peak appears in the combined data, however the structure has become very doubled peaked.  No periodic variations were found in the MACHO or OGLE II data.  This optical period, which we interpret as the orbital period, places this HMXB well within the limits of the proposed Be/X-ray binaries on the Corbet diagram \citep{coe08,cor99}.  The ephemeris of the outbursts is MJD $(53178.3\pm0.8) + n(17.79\pm0.01)$. 

\subsection{{\it RXTE} Observations}
Reviews of our ongoing {\it RXTE} X-ray monitoring program of the SMC may be found in \citet{gal08} and \citet{laycock05}.  During 2006 a trial observation at Position 5 (12.5$^\circ$, -73.1$^\circ$) revealed two pulsars to be in outburst, SXP18.3 and SXP15.3.  From 3 August 2006 (MJD\,53950) till 19 June 2007, this position was monitored weekly.  After three weeks of monitoring on 30 August 2006 (MJD\,53977) SXP18.3 switched back on.  The source remained in outburst and was continually detected every week for the next 36 weeks finally disappearing on 3 May 2007 (MJD\,54223).  This was the longest uninterrupted outburst ever seen in the SMC. Figure \ref{fig:xlc} shows the full 10 years of {\it RXTE} monitoring, the collimator response represents the position of the source within the field of view (FoV) of RXTE (1 being at the centre and 0 the edge).  The red points represent where 18.3\,s pulsations above a 99\% local significance level have been detected in the power spectra of {\it RXTE} light curves.  The blue points represent no detection. For specific details of the data processing see \citet{gal08}.  As can be seen prior to the massive outburst the source was only seen in outburst on seven occasions.  These outbursts do not correlate to any particularly interesting features on the optical light curve.

We have estimated the range of luminosities for this outburst using PIMMS\,v3.9f\footnote[1]{(http://heasarc.gsfc.nasa.gov/Tools/w3pimms.html)} and assumed a distance to the SMC of 60\,kpc.  We used the spectral fit found from the {\it XMM-Newton} observations \citep{eger08} ($\Gamma = 0.65$, N$_{H} = 2\times10^{22}$cm$^{-2}$) and assumed that the {\it XMM-Newton} measured pulse fraction of $21\pm3$\% remained constant for the entire observation.   The {\it RXTE} luminosity range throughout the outburst is $L_{0.2-10}=(0.51-3.2)\times10^{37} erg/s$.

\begin{figure}
 \includegraphics[width=60mm, angle=90]{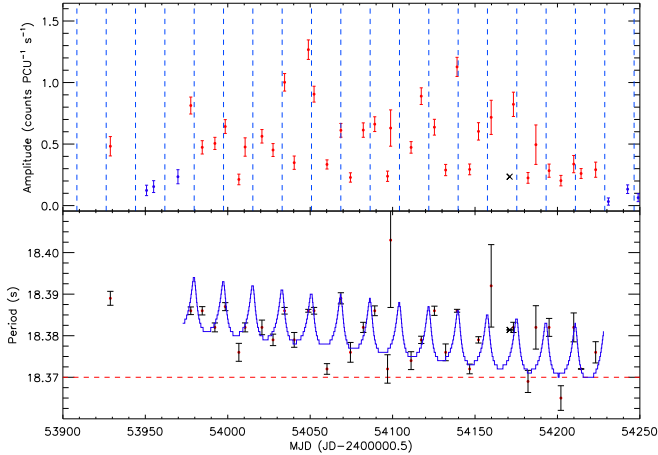}
  \caption{SXP18.3 X-ray outburst. Top panel shows the {\it RXTE} amplitude (red points are above 99\% significance) with the vertical dotted lines showing periastron passage as predicted by the optical light curve.  The bottom panels shows the detected {\it RXTE} pulse periods with the orbital solution over plotted. The black crosses are the {\it XMM Newton} detected pulse period (orbital phase 0.80) and flux converted to {\it RXTE} with PIMMS\,v3.9f. \label{fig:outburst}}
\end{figure}

\begin{table}                                                   
 \begin{minipage}{80mm}                                             
  \caption{Orbital parameters \label{ta:param}}                      
  \begin{tabular}{@{}lr@{}}                                         
  \hline                                                             
   Parameter & Orbital Solution \\                                   
 \hline                                                              
 P$_{orbital}$ (d)            &$ 17.79_{fixed}              $\\                               
 a$_{x}$sin{\it i} (light-s)  &$ 75\pm3                     $\\                                
 $\omega$ ($^{o}$)            &$ 15\pm6                     $\\                                
 e                            &$ 0.43\pm0.03                $\\                               
 $\tau_{periastron}$ (MJD)    &$ 53997.6\pm0.2              $\\                             
 P$_{spin}$ (s)               &$ 18.3854\pm0.0004           $\\                                
$\dot{P}$ (s/yr)              &$ [-1.79\pm0.11]\times10^{-2}  $\\  
 $\chi^{2}_{\nu}$             & 8.22 \\         
 \hline                                                              
\end{tabular}\\                                                      
\end{minipage}                                                       
\end{table}     

\subsubsection{Orbital fitting}
Figure \ref{fig:outburst} shows in detail the detected luminosities and pulse periods throughout the 9 month long X-ray outburst.  It is evident from the general trend in the detected pulse period (bottom panel) that SXP18.3 was spinning up during the outburst. This general spin up equates to a luminosity of $L_{3-10}=2.67\times10^{36} erg/s$ \citep{ghosh79}.  The shorter variations occurring throughout the outburst suggest that the detected spin period is being orbitally modulated.  We detrended the data and performed a Lomb-Scargle period search revealing a peak in the power spectrum above the 99\% significance level at $17.62\pm0.1\,d$.  This period would seem to be consistent to that found in the optical data.  We have thus tried to fit a full orbital solution to the pulse period values.  The orbital fitting is performed by a chi-squared minimisation method.  We performed two fits, the first allowing all the parameters to vary, and in the second we fixed the orbital period to that found from the optical light curve.  Subsequent folds of the {\it RXTE} light curve using the first fit appeard quite chaotic and unbelieveable, hence the model is not presented. In Figure \ref{fig:outburst} we show the orbital fit to the data with the orbital parameters shown in Table \ref{ta:param}.  It is worth noting that the time of periastron passage is consistent with that from the independent analysis of the optical data.  The orbital parameters are typical of those seen in other accreting Be/X-ray binaries \citep{oka01}, most Be systems have P$_{orbital}\sim$20-100\,d with eccentricities in the range 0.3-0.5 \citep{bild97}.  
                                                    
\begin{figure}
 \includegraphics[width=60mm, angle=0]{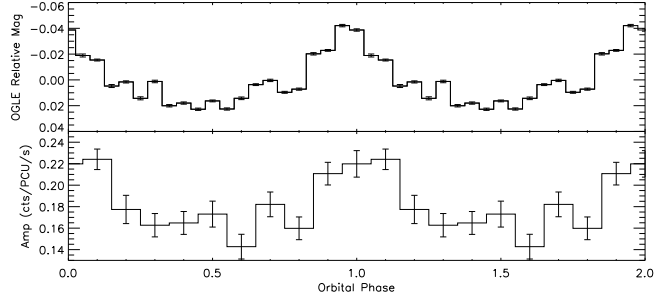}
  \caption{{\it RXTE} (bottom) and OGLE (top) folded light curves.\label{fig:folded_lc}}
\end{figure}

We have folded the first two years of the detrended OGLE III light curve and the {\it RXTE} amplitude measurements to evaluate the orbital solution.  The {\it RXTE} data are folded in 10 bins and the OGLE III data have been folded in 20 bins using the orbital period and ephemeris from Table \ref{ta:param}.  Both the X-ray and optical folded light curves (bottom and top respectively of Figure \ref{fig:folded_lc}) have fairly sinusoidal profiles and are well matched with the peaks coinciding.  This would suggest that 17.79\,d is the orbital period. There is also another probable peak in the optical data at phase 0.7 with some hint of its presence in the X-ray fold. 

\subsubsection{{\it RXTE} spectra and pulse profiles}
Because {\it RXTE} has such a large field of view we constantly cover many SMC sources in our observations.  This makes it extremely difficult to resolve one source from another.  Due to the unusually large outburst from SXP18.3 there were several observations where our analysis revealed that it was the only known source to be in outburst.  We placed further restrictions on which observations were suitable for analysis by only examining those observations when the source was detected above the 99\% global significance level \citep[for details see][]{gal08}.  We examined the spectra of 18 out of the 35 observations in the 3-10keV range.  Although SXP18.3's outburst was very long it was never particularly bright, hence the spectra have a low signal to noise. We attempted to fit each spectrum with an absorbed power law and tried both fixing the absorption to that of the SMC and allowing it to vary.  We were unable to produce any meaningful fits to the spectra.  This is most likely due to the unknown contamination from background AGN, and X-ray binaries emitting at a very low level.

\begin{figure}
 \includegraphics[width=85mm, angle=0]{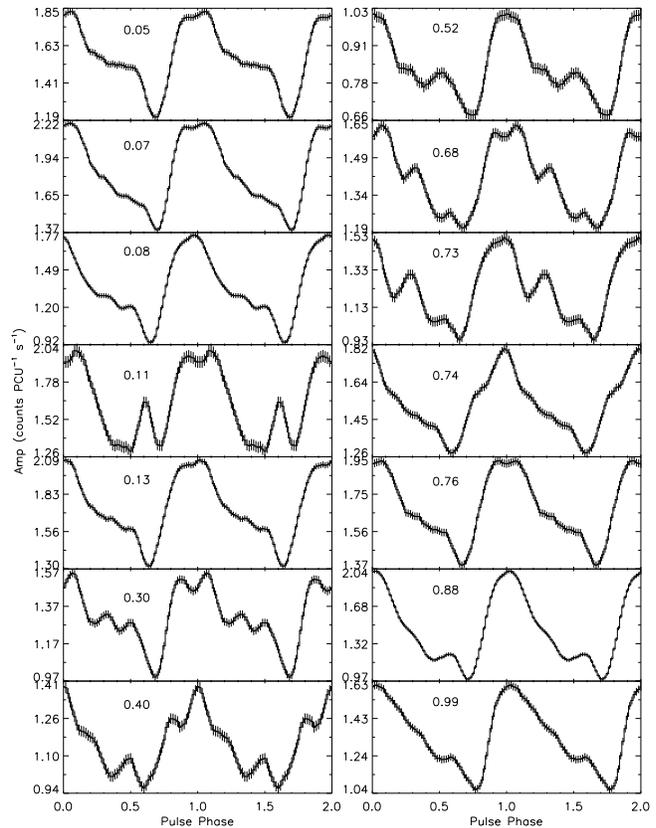}
  \caption{Folded pulse profiles of SXP18.3, an arbitrary pulse phase shift was applied to align the peaks since they cannot be phase locked between observations.  Arranged in order of orbital phase (given in the label) starting at the top left.\label{fig:pulse_profile_all}}
\end{figure}

In addition to analysing the spectra we have also examined the pulse profiles for each of these observations. We used the ephemeris and orbital period from Table \ref{ta:param} to calculate the orbital phase for each observations (shown on each pulse profile), the observations were then arranged to appear in order of their orbital phase, duplicated orbital phases were removed (Figure \ref{fig:pulse_profile_all}). In each case an arbitrary pulse phase shift was applied to align the main peak of the profile with phase 0.  The majority of the pulse profiles present a very similar sinusoidal shape with a very steep rise and gradual fall, in good general agreement with those presented in \citet{haberl08}.  The pulse profiles around periastron passage appear to be fairly stable with small changes to the shape of the fall off. There is some emission that is associated with pulse phase 0.5 with a subsequent steep decline to minimum emission occurring at pulse phase 0.7-0.8.  This emission is possibly due to either the X-ray beam fanning out and becoming more conical in shape allowing emission from the second pole to be seen, or that the accretion stream is preferential to one pole and only occasionally splits to both poles.  When the neutron star is around apastron (orbital phase 0.3-0.7) the pulse profiles becomes rather more complicated with several additional peaks emerging.   We have used these pulse profiles to measure the pulsed fraction to be in the range $22\pm10$\%.  This is consistent with the value found from the {\it XMM-Newton} observations.  There is also a strong correlation between the pulse fraction and the observed amplitude, suggesting that although the source would appear to vary throughout the outburst (Figure \ref{fig:outburst}) it is infact emitting at an almost constant stable rate.

\section{Discussion}
The optical and X-ray data suggest that SXP18.3 is a HMXB transient.  From analysis of the MCPS data we have proposed a tentative classification of of the counterpart as B2V, supporting the proposal by \citet{eger08} that SXP18.3 is a Be/X-ray binary.   The optical light curve of SXP18.3 shows periods of extreme brightening occurring over very long time scales.  These dramatic outbursts are seen to occur in many other Be/X-ray binary systems \citep[e.g. SXP46.6, SXP6.85][]{kem08} and are most likely due to massive size variations in the circumstellar disk surrounding the Be star. In particular SXP6.85 \citep{kem08} exhibits similar recurrent optical outbursts of comparable duration and brightness.  Complex computer simulations are required to understand the exact nature of these optical outbursts and how they produce the colour variations.  The 17.79\,d period found in the optical data is interpreted as the orbital period of the neutron star round the Be star.  If we assume that the equatorial plane of the Be star is approximately coincident with that of the neutron stars orbit, then as periastron is approached we see some distortion of the circumstellar disk due to the larger gravitational attraction between the two objects \citep{oka01}.  These distortions would temporarily increase both the size and luminosity of the disk.  This orbital modulation is seen in many Be/X-ray binary systems \citep{coe08}. 
 
It is interesting that the orbital modulation is not always present, only being significantly detected when the I magnitude of the Be star is brighter than 15.2\,mag.  This suggests that the disk has to be sufficiently large and hence close to the neutron stars orbit for any significant perturbations.  However, we note that during the massive X-ray outburst there is no modulation in the optical data. This is possibly due to the extreme nature of this outburst. Here we may have a situation where the disk has possibly grown to such a size that it is encompassing the entire orbit of the neutron star, allowing continuous accretion. Assuming typical values for the mass of the Be and neutron star and the orbital period found, we would expect a semi-major axis of $\sim$85R$_{\odot}$.  From H$\alpha$ equivalent width measurements \citet{grund07} have estimated the disk radius in the Be/X-ray binary A0535+26 to reach a maximum size of 90R$_{\odot}$.  It is not unreasonable to envisage a disk growing to this size and producing such a massive outburst.  If this situation is correct then the perturbation suffered by the disk may increase the local density of material rather than extend the physical size of the disk, resulting in the orbital signature disappearing. One problem with this interpretation is explaining the orbital signature seen in the first two years of detrended OGLE III data.  If the optical outburst is attributed to the disk growing in size then during these two years the disk is larger or denser than when we observed the X-ray outburst.  If the model proposed is correct then the neutron star would have been within the disk for this entire period and we should have seen no optical signature and possibly even X-ray outbursts.  Unfortunately, for the majority of this period SXP18.3 was outside the central $1^{\circ}$ of {\it RXTE}'s FoV, thus having a much lower detection sensitivity. SXP18.3 may have dropped below this threshold and hence we cannot say for definite that SXP18.3 was truly off.  It is interesting to note that prior to the massive X-ray outburst we only detected SXP18.3 on a handful of occasions, all during periods when I$<$15.2mag.

We attempted to fit a full orbital solution to the variations in the observed spin period.  Our foremost problem with this fit is that the data sampling is too infrequent to allow us to take into account the variations in the accretion torques.  This will manifest its self as deviations away from our orbital solutions.  Even though we have data spanning many orbital cycles during the outburst the fact that we only have two or three data points per orbital cycle is simply insufficient to properly constrain the orbital parameters. However, the close match between the X-ray and optical folded light curves clearly suggests that 17.79\,d is the true orbital period.  

SXP18.3 lies in the bottom left corner in the low orbital and spin period regime of the Corbet diagram.  It is right on the extremity of the distribution of Be/X-ray binaries.  If this source was to move slightly further to the left (lower orbital period) then it would start to fall into the Roche lobe overflow group.  When we examine the type of outbursts exhibited by SXP18.3 we see that there are a couple of short lived typical Type I outbursts just above our detection threshold, and then the massive 36 week Type II outburst.  Outbursts of this length are extremely rare for Be/X-ray binaries, since for the duration of this period the neutron star must be accreting constantly.  SAX J2103.5+4545 is a Be/X-ray binary where similar X-ray behaviour is seen \citep{Baykal02}.  It is a pulsar with a long spin period of $\sim$359\,s but a very short orbital period of 12.7\,d (comparable with SXP18.3).  This would place it at the top left of the distribution of Be/X-ray binaries in the Corbet diagram. SAX J2103.5+4545 has exhibited several bright X-ray phases lasting hundreds of days, with luminosities $\sim10^{36}$\,erg\,s$^{-1}$.  It is only during these bright phases when the orbital period has been detected.  \citet{Blay04} suggest that this could be due to the density/size of the circumstellar disk being large enough to temporally fill its Roche lobe during periastron, and hence temporally allow accretion to take place.  It could be that the moderately bright outburst seen in SXP18.3 is due to a very similar mechanism to that which produces the outbursts seen in SAX J2103.5+4545.  We could imagine that these short orbital period systems possibly flip between states of Roche lobe overflow and quiescence.  Monitoring the state of the H$\alpha$ line before, during and after an outburst would help shed some light on these systems.  Be/X-ray binary systems like these may well be providing an indication of a lower limit for Be/X-ray binaries on the Corbet diagram.

\section{Conclusions}
In light of the optical and X-ray data we feel that SXP18.3 is the newest member of the group of transient Be/X-ray binaries in the SMC.  The proposed orbital model requires intensive observations and detailed computer simulations to fully assess the parameters and explain the presence of the second optical peak.  We are now beginning to explore the limits of the Corbet diagram where mixtures of behaviours are observed and sources like SXP18.3 are key to this study.  

\section*{Acknowledgements}
This publication makes use of data products from the Two Micron All Sky Survey, which is a joint project of the University of Massachusetts and the Infrared Processing and Analysis Center/California Institute of Technology, funded by the National Aeronautics and Space Administration and the National Science Foundation. This paper utilises public domain data originally obtained by the MACHO Project, whose work was performed under the joint auspices of the U.S. Department of Energy, National Nuclear Security Administration by the University of California, Lawrence Livermore National Laboratory under contract No. W-7405-Eng-48, the National Science Foundation through the Center for Particle Astrophysics of the University of California under cooperative agreement AST-8809616, and the Mount Stromlo and Siding Spring Observatory, part of the Australian National University.  This research has made use of the SIMBAD data base, operated by CDS, Strasbourg, France.  We acknowledge support from the NASA {\it Swift} team in providing ToO observations in order to locate the source.  The OGLE project is partially supported by the Polish MNiSW grant N20303032/4275.

\bsp

\label{lastpage}

\end{document}